\documentclass[12pt,preprint]{aastex}

\begin{document}

\title {THE VERY SLOW WIND FROM THE PULSATING SEMIREGULAR RED GIANT,  L$_{2}$ PUP}

\author{M. Jura, C. Chen and P. Plavchan} 
\affil{Department of Physics and Astronomy, University of California,
    Los Angeles CA 90095-1562; jura@clotho.astro.ucla.edu; cchen@astro.ucla.edu; plavchan@astro.ucla.edu}

\begin{abstract}
We have obtained 11.7 ${\mu}$m and 17.9 ${\mu}$m images at the Keck I telescope of the circumstellar dust emission from L$_{2}$ Pup, one of the nearest ($D$ = 61 pc) mass-losing, pulsating, red giants 
that has  a substantial infrared excess.  We propose that the star is losing mass at a rate of ${\sim}$3 ${\times}$ 10$^{-7}$ M$_{\odot}$ yr$^{-1}$.    Given its relatively low luminosity (${\sim}$ 1500 L$_{\odot}$), relatively high effective temperature (near 3400 K),  relatively short period (${\sim}$ 140 days), and the inferred gas outflow speed of 3.5 km s$^{-1}$, standard models for dust-driven mass loss do not apply.  Instead, the wind may be driven by the stellar pulsations with radiation pressure on dust being relatively unimportant,
as described in some recent calculations. L$_{2}$ Pup may serve as the prototype of this phase of stellar evolution where 
it could lose ${\sim}$15\% of its initial main sequence mass.    
\end{abstract}

\section{INTRODUCTION}

Red giants with luminosities near  10$^{4}$ L$_{\odot}$ typically display infrared excesses that are
the result of dust formed in a stellar wind reprocessing optical light emitted by the central star (Habing 1996).  The mass loss rates can exceed 10$^{-5}$ M$_{\odot}$ yr$^{-1}$ and the outflow speed is typically near 15 km s$^{-1}$.  These winds are important in stellar evolution because the star may lose over half of its initial main sequence mass during this phase.  

A standard model to explain this mass loss is that as the stars pulsate, shock waves form in the atmosphere which drive matter above the nominal photosphere.
In the post-shock region,  the matter cools and  grains form.  After the dust is created, 
 radiation pressure then expels the matter into the interstellar medium (see, for example, Lamers \& Cassinelli 1999, Willson 2000).  Although the standard model successfully explains the properties of many red giants winds, there are some systems where
it does not apply (see Jura \& Kahane 1999).  This ``standard model" does not
consider the effects of binaries, rotation or magnetic fields.

In a recent set of  calculations of models for mass loss from  pulsating red giants, Winters et al. (2000) found  a class of solutions, that are distinct from the standard model and which the authors denote as ``case B", where the mass loss rate is less than 3 ${\times}$ 10$^{-7}$ M$_{\odot}$ yr$^{-1}$ at an outflow speed less than 5 km s$^{-1}$.  In these models,  mechanical energy from the pulsations drives the mass loss; radiation pressure on grains is a relatively minor effect.  These ``case B" models 
apply to stars of relatively low luminosity, typically where $L_{*}/M_{*}$ $<$
3500 $L_{\odot}/M_{\odot}$, the pulsational periods are less than 300 days and the effective temperatures of the mass-losing stars are 
greater than 2600 K.  

 Winters et al. (2000) were not able to identify any stars which clearly exhibit ``B-class" mass loss.  Here, we report
detailed mid-infrared observations obtained at the 10m Keck telescope of L$_{2}$ Pup, a relatively nearby 
pulsating, mass-losing red giant.    We propose that our data for the circumstellar shell around L$_{2}$ Pup can be
best interpreted as a ``B-class" wind.  

\section{L$_{2}$ Pup}
L$_{2}$ Pup (= HD 56096 = HR 2748) has a distance measured by the {\em Hipparcos} satellite of 61 ${\pm}$ 5 pc, thus making it, along with R Dor, one of the  pulsating,  mass-losing red giants that is nearest to the Sun.    In the Yale Bright Star Catalog it is
classified as M5IIIe and assigned a visual magnitude of 5.10. It is a semiregular variable with a period of
140.6 days according to the General Catalog of Variable Stars, but with a 137 day period according to Whitelock et al. (2000).

With a time-averaged measure that m$_{K}$ = - 2.24 mag and thus m$_{bol}$ = 0.73 mag (Whitelock et al. 2000), and  if $M_{Bol}(Sun)$ = +4.72 mag, then the  luminosity of L$_{2}$ Pup is
1500 L$_{\odot}$.  The time-averaged value of (J-K) is 1.31 mag (Whitelock et al. 2000), and therefore,  from the models presented by Bessell et al. (1989) and Houdashelt et al. (2000), the time-averaged effective temperature of L$_{2}$ Pup is about
3400 K.           
 With these values of the luminosity and temperature, the time-averaged radius of L$_{2}$ Pup is 7.8 ${\times}$ 10$^{12}$ cm.

Since L$_{2}$ Pup has a velocity relative
to the Galactic Plane of ${\sim}$90 km s$^{-1}$ (Feast \& Whitelock 2000), it probably  belongs to the thick disk population, and thus  the mass of the main sequence progenitor of the star was about 1 M$_{\odot}$ (Jura 1994).  L$_{2}$ Pup's location on the H-R diagram can be reproduced by interpolating from the   Lattanzio's (1991) models for the Asymptotic Red Giant Branch evolution of a star with a metallicity that is 
0.5 solar.  In these calculations, the star has  a luminosity between 1000 L$_{\odot}$ and 1500 L$_{\odot}$ for about 5 ${\times}$ 10$^{5}$ yr.  Given its luminosity and effective temperature, it is also possible that L$_{2}$ Pup is a first ascent red giant (see the tracks by Girardi et al. 2000);
the luminosity and color, do not uniquely indicate the star's evolutionary phase.  The detection of Tc would indicate that the star almost certainly lies on the AGB; Lebzelter \& Hron (1999) report a ``possible" detection of Tc
in the atmosphere of L$_{2}$ Pup.

L$_{2}$ Pup is  a visual binary with a
companion which in the year 1913 was at 62{\arcsec} separation (Proust, Ochsenbein \& Pettersen 1981).  However, in the {\em Hipparcos} data,
this companion now lies at an angular separation of 101{\arcsec} (Dommanget \& Nys 2000).  This much larger separation in the {\em Hipparcos} data implies
that the two stars are not a physical pair.

Evidence that L$_{2}$ Pup is losing mass comes from its infrared excess, the detection of circumstellar molecules, and its large, time-varying optical polarization.  The value of F$_{\nu}$(12 ${\mu}$m)/F$_{\nu}$ 2.2 (${\mu}$m) is typically near 0.40 to 0.50, which is much larger than 0.07, the characteristic photospheric value for this ratio of M giants (see Jura, Webb \& Kahane 2001).  Since the  circumstellar shell
dominates the emission from the system for ${\lambda}$ ${\geq}$ 5 ${\mu}$m, 
the dust shell re-emits approximately 10\% of the total stellar luminosity.   

Detected circumstellar  molecules include H$_{2}$O, OH, SiO and CO. The line profiles of these molecules differ from each other because the H$_{2}$O, OH, and SiO lines are produced as masers while the CO is thermally excited.    The H$_{2}$O profile is variable with a velocity peak at $V_{LSR}$ = 38 km s$^{-1}$ and a second component that shows emission at velocities as low as  $V_{LSR}$ = 31 km s$^{-1}$ (Lepine, Paes de Barros \& Gammon 1976, Gomez Balboa \& Lepine 1986).  The OH maser emission has been detected at $V_{LSR}$ = 33-34 km s$^{-1}$ (Dickinson et al. 1986).  The SiO  maser emission is variable and has been detected at $V_{LSR}$ = 43 km s$^{-1}$ (Balister et al. 1977) and as a triple-peaked source with velocities between $V_{LSR}$ = 32 km s$^{-1}$ and 42 km s$^{-1}$ (Haikala 1990).  The CO emission peaks for the different rotational transitions have been reported between $V_{LSR}$ = 32.9 km s$^{-1}$ and 33.7 km s$^{-1}$ (Kerschbaum et al. 1996, Kerschbaum \& Olofsson 1999).  The outflow velocity derived from the full extent of the CO emission has been measured to be between 2.0 km s$^{-1}$ and 3.3 km s$^{-1}$.    

The SiO maser emission typically arise within 1-2 stellar radii of the photosphere (see Doeleman, Lonsdale \& Greenhill 1998) and thus measures the  motions induced by the stellar pulsations rather than the gas outflow. The  velocities of the true circumstellar molecules range from 
31 km s$^{-1}$ to 38 km s$^{-1}$;  the midpoint of this range of circumstellar gas velocities is 34.5 km s$^{-1}$.  The extreme values of the circumstellar gas emission are determined from the H$_{2}$O masers, but the CO lines are nearly
as wide as this spread in the H$_{2}$O radial velocities.  
Feast \& Whitelock report that  optical measurements of the star's
heliocentric radial velocity  yield 53.0 km s$^{-1}$ which corresponds to $V_{LSR}$ = 34.2 km s$^{-1}$.  Within the uncertainties, it appears that the center of mass velocity of L$_{2}$ Pup is 34.5 km s$^{-1}$ and that circumstellar molecules move with speeds ${\pm}$ 3.5 km s$^{-1}$ with respect to this central velocity.  Although the outflow is not spherical, we have no information on
the spatial variations of the gas outflow speed and we 
  therefore
assign an outflow speed of 3.5 km s$^{-1}$ to the wind.  This wind speed 
is much lower than the typical value of ${\geq}$10 km s$^{-1}$ found for most mass-losing red giant stars (Loup et al. 1993).  To illustrate the point that L$_{2}$ Pup has an unusually low wind velocity, we show in Figure 1 a histogram of the outflow velocities measured from the line widths of the circumstellar CO emission from a survey of semiregular and irregular variables (Kerschbaum \& Olofsson 1999).  In this survey, L$_{2}$ Pup has the narrowest line, although there are some other stars such as EP Aqr with narrow spikes on top of a broader profile (see also Knapp, Young, Lee \& Jorissen 1998).    

At times, the optical light from L$_{2}$ Pup is over 10\% polarized (Magalhaes et al. 1986). 
Since L$_{2}$ Pup is only 61 pc from the Sun, this large polarization cannot be produced by interstellar grains but must be intrinsic to
the star and its circumstellar envelope.  The position angle varies with time and has ranged from 156$^{\circ}$ through 0$^{\circ}$ up to 60$^{\circ}$ (Magalhaes et al. 1986).  The large optical polarization implies
that the circumstellar envelope is anisotropic.

\section{OBSERVATIONS}

Our data were obtained on  2001 Feb 05 (UT) at the Keck I telescope 
using the Long Wavelength Spectrometer (LWS) which was built by a team 
led by B. Jones and is described on the Keck web page.  The LWS is a 
128 ${\times}$ 128 SiAs BIB array with
a pixel scale at the Keck telescope of 0{\farcs}08 and a total field of 
view of 10{\farcs}2 ${\times}$ 10{\farcs}2.  We used the ``chop-nod" 
mode of observing, and 2 different filters centered at 11.7 ${\mu}$m and 17.9 ${\mu}$m with widths of 1.0 ${\mu}$m and 2.0 ${\mu}$m, respectively.  Following Chen \& Jura (2001), we used Capella (= HR 1708)  for flux and point-spread-function calibrations.  For Capella the FWHM of the image was 0{\farcs}47 and 0{\farcs}49 at 11.7 ${\mu}$m and 17.9 ${\mu}$m,  respectively.   

The images of L$_{2}$Pup in the 11.7 ${\mu}$m in the 17.9 ${\mu}$m  filters are presented in 
Figures 2 and 3.  The 11.7 ${\mu}$m image of Capella is shown in Figure 4. 
Because the calibrator was much more compact than L$_{2}$ Pup,  deconvolution
of the data is not required for offsets greater than 0{\farcs}6 from the star
since at these offsets, at most azimuths, the central point source  contributes less than 20\% of the observed intensity.   In the mid-infrared,  L$_{2}$ Pup is  brighter  than Capella, and we cannot be sure that some of the very faint extended emission that we detected
is real rather than an artifact.  We therefore only show
emission at positions where the observed intensity is at least 2\% of the peak intensity.  
At both wavelengths, we see an extended morphology oriented along an axis at position angle 135$^{\circ}$ with an additional  distinct blob
at position angle of 225$^{\circ}$.    

Integrated over the image, we measure F$_{\nu}$(11.7 ${\mu}$m) = 2500 Jy; this is comparable to the non-color corrected value of F$_{\nu}$(12 ${\mu}$m) = 2400 Jy measured with IRAS and the results from DIRBE (see Figure 7 below)
that F$_{\nu}$(12 ${\mu}$m) ranges between about 2000 Jy and 2200 Jy.   

If we assume the same atmospheric extinction for L$_{2}$ Pup and Capella, our data imply that F$_{\nu}$(17.9 ${\mu}$m) = 1100 Jy. However, because of its southern declination, there was an air mass of 2.32 during our  observations of L$_{2}$ Pup. Therefore, the atmospheric extinction correction for the flux at 17.9 ${\mu}$m of L$_{2}$ Pup was  substantial and the absolute level of our measurements at 17.9 ${\mu}$m is uncertain.  The data from the DIRBE satellite shown below in Figure 7 shows
that F$_{\nu}$(12 ${\mu}$m)/F$_{\nu}$(25 ${\mu}$m) remains constant to within
5\% over a pulsational cycle.  Interpolating the DIRBE data between 12 ${\mu}$m and 25 ${\mu}$m indicates an average flux at 17.9 ${\mu}$m of 1700 Jy instead of our measured value of 1100 Jy.  We therefore apply an uncertain correction factor of 1.5 to our measured values of the intensities at 17.9 ${\mu}$m.

\section{MODELS}
To model the infrared maps of the envelope, we  follow a standard prescription
where the dust grains are heated
by light from the central star and then re-radiate in the infrared (Sopka et al. 1984).   Such models require estimates for the size and composition of the grains. Here, we use simplified semi-analytic models to show  how 
different assumptions and parameters propagate into our derived results.  

\subsection{Grain Properties}
An important parameter is the grain size.  
 Daniel
(1982) has modeled the optical polarization around L$_{2}$ Pup by spherical grains with
radius $a$ = 0.26 ${\mu}$m with an assumed simple bipolar
configuration of the circumstellar envelope.  In fact, as shown by our infrared imaging,  there is a blob at position angle 225$^{\circ}$ in addition to the extended bipolar morphology extended along an axis at position angle 135$^{\circ}$.  The spatial structure of the circumstellar dust therefore is more complex than presumed in Daniel's models. Everything else being equal, the more complex geometry will
result in less integrated polarization.  Therefore,  in order to reproduce
the result that the polarization in the integrated light approaches 10\%, the grains must be more efficient at producing polarization and therefore smaller than those postulated by Daniel.  Jura (1996) suggested that the typical spherical particle around a mass-losing oxygen-rich star has a radius smaller than 0.15 ${\mu}$m. 
Here, we assume that most of the particles around L$_{2}$ Pup are similar to those
around other oxygen-rich stars and thus for most of the particles $a$ ${\leq}$ 0.1 ${\mu}$m.  
Since $T_{*}$ = 3400 K,     this estimate of the grain size is such that $a$ ${\leq}$ ${\lambda}$/(2${\pi}$) for all wavelengths of interest both for emission in the infrared and for absorption of the light from the photosphere.  In this case, the temperature
of the grains is insensitive to their size.

 A model of  the grain emissivity is necessary to explain our infrared data.  Unfortunately,  
the spectrometer on the IRAS satellite  did not acquire a good infrared spectrum of  this object (Sloan \& Price 1998, Volk \& Cohen  1989), and  ground-based infrared spectra have not been reported in the literature surveyed with SIMBAD.  Therefore, the magnitude of silicate features in the infrared dust emission cannot be easily quantified.    
 If we define $Q_{abs}$ as the ratio of the absorption cross section to the geometric cross section (see, for example, Spitzer 1978),  then the relationship between the opacity, ${\chi}_{\nu}$ (cm$^{2}$ g$^{-1}$) and $Q_{abs}$ is
that ${\chi}_{\nu}$ = $(3\,Q_{\nu})/(4\,{\rho}_{s}\,a)$ where ${\rho}_{s}$ (g cm$^{-3}$) is the density of the solid grains which we take to equal
3.3 g cm$^{-3}$ (Kim, Martin \& Hendry 1994).  A simplified version of
the circumstellar 
silicates described by Ossenkopf, Henning \& Mathis (1992) and David \& Pegourie (1995) can be expressed as:
$Q_{abs}$ = $K_{1}\,{\nu}^{+1}$ for ${\lambda}$ ${\leq}$ 9 ${\mu}$m and
$Q_{abs}$ = $K_{2}\,{\nu}^{+1}$ for ${\lambda}$ ${>}$ 9 ${\mu}$m  
 with $K_{1}/K_{2}$ ${\approx}$ 0.2.  This simplified model for $Q_{\nu}$ implies values of ${\chi}_{\nu}$ from which the grain temperature can be inferred, although  this model does not include any spectral features such as
those produced by  silicates.

Given ${\chi}_{\nu}$, then the grain temperature profile can be found
following the prescription that the particles are heated by starlight and
re-radiate in the infrared.  
For  L$_{2}$ Pup  with $T_{eff}$ = 3400 K, then typically the heating occurs near ${\lambda}$ ${\sim}$ 1.5 ${\mu}$m, while  most of the re-emission in the  infrared occurs
for  ${\lambda}$ $>$ 9 ${\mu}$m.   
Therefore, following Sopka et al. (1984),  the grain temperature, $T_{gr}$, as a function of distance, $R$, from the star is given as:
\begin{equation}
T_{gr}\;=\;(\frac{K_{1}}{4\,K_{2}})^{0.2}\,T_{*}\,(\frac{R_{*}}{R})^{0.4}
\end{equation}  
A validation of this approach is to compare our simple model with more sophisticated calculations by other authors.  For example,  for a star with $T_{eff}$ = 3400 K, at $D$ = 20 $R_{*}$, equation (1) yields $T_{gr}$ = 563 K.  
With detailed calculations for small silicate grains,   Lorenz-Martins \& Pompeia (2000) 
compute that at $R$ = 20 $R_{*}$, $T_{gr}$ = 567 K.   The closeness of this agreement is fortuitous, but it does allow that our simple model might
be a useful approximation.

We estimate the intensity of the circumstellar emission with the assumptions that the mass loss rate, ${\dot M}_{gr}$, and the grain outflow velocity, $V_{gr}$, are constant.    
If we observe along a ray where the measured  distance is denoted by $z$, then
 the observed surface brightness, $I_{\nu}$, (Jy ster$^{-1}$) is: 
\begin{equation}
I_{\nu}\;=\;{\int}\,{\chi}_{\nu}\,{\rho}(z)\,B_{\nu}(T(z))\,dz
\end{equation}
For a spherically symmetric distribution, 
\begin{equation}
{\rho}(z)\;=\;\frac{\dot{M}_{gr}}{4\,{\pi}\,(b^{2}\,+\,z^{2})\,V_{gr}}
\end{equation}
where $b$ denotes the  impact
parameter of the sight-line with respect to the central illuminating star. The solution to equations (2) and (3)  for ${\dot {M}_{gr}}$ is:  
\begin{equation}
{\dot{M}_{gr}}\;=\;\frac{4\,b\,V_{gr}\,I_{\nu}\,f_{corr}({\nu})}{{\chi}_{\nu}\,B_{\nu}[T_{gr}(b)]}
\end{equation}
The function $f_{corr}({\nu})$ is a measure of the effects of a non-uniform temperature distribution of the dust grains and must be computed numerically.  If the grains had a uniform temperature, then $f_{corr}({\nu})$ = 1.  
For L$_{2}$ Pup, we estimate that at 1${\arcsec}$ from the star that  $T_{gr}$ = 280 K.  In this case, $f_{corr}$(11.7 ${\mu}m$) = 2.0 while $f_{corr}$(17.9 ${\mu}$m) =
1.7.

If ${\chi}_{\nu}$ is known, then we can invert equation (4) to find
from our measurements of I$_{\nu}$, the value of $T_{gr}$ as a function of $b$.  Assuming some silicate emission, we expect  from Draine (1985) and Ossenkopf et al. (1992) 
that ${\chi}_{\nu}$(11.7 ${\mu}$m)/${\chi}_{\nu}$(17.9 ${\mu}$m) = 1.6, instead of 1.5 as predicted if ${\chi}_{\nu}$ ${\propto}$ ${\nu}$. 
With this ratio of the opacities and with the
measurement that $I_{\nu}$(11.7 ${\mu}$m)/$I_{\nu}$(17.9 ${\mu}$m) = 2 ${\pm}$ 0.7 at
offset 1${\arcsec}$ from the star, we derive that at this location,  
T$_{gr}$ ${\geq}$ 350 K.  From equation (1), we predict that $T_{gr}$ = 280 K.  Given the uncertainty in the calibration of the intensity at 17.9 ${\mu}$m,  and our lack of constraints on the grain composition, the
simplified model and the data are consistent with each other.  

To estimate ${\dot {M}_{gr}}$, we need to infer the absolute value of ${\chi}_{\nu}$ as well as its variation as a function of frequency.  
From Ossenkopf et al. (1992),  we adopt ${\chi}_{\nu}$(11.7 ${\mu}$m) = 2000 cm$^{2}$ g$^{-1}$.  In the simplified model for the particle emissivity described above, at the peak of the photospheric emission,
 ${\chi}_{\nu}$(1.5 ${\mu}$m) ${\sim}$ 3000  cm$^{2}$ g$^{-1}$.   

\subsection{Mass Loss Rate}
To estimate the mass loss rate from equation (4), we  must determine $V_{gr}$.  In the envelopes of red giants the grains stream supersonically through the gas, and
the grain velocity is therefore larger than the gas outflow velocity, $V_{ga}$.  If the gas outflow rate is ${\dot{M}_{ga}}$, and if $a$ $<$ ${\lambda}$/(2${\pi}$) so that scattering by the grains is unimportant, then
 (Lamers \& Cassinelli 1999)
\begin{equation}
V_{gr}\;=\;V_{ga}\;+\;(\frac{Q_{abs}L_{*}\,V_{ga}}{{\dot M}_{ga}c})^{1/2}
\end{equation}
Assuming the emission from the star with $T_{eff}$ = 3400 K  peaks near 1.5 ${\mu}$m, and 
with ${\chi}_{\nu}$ (1.5 ${\mu}$m) = 3000 cm$^{2}$ g$^{-1}$, $V_{ga}$ = 3.5 km s$^{-1}$, ${\dot M}_{ga}$ = 3 ${\times}$ 10$^{-7}$M$_{\odot}$ yr$^{-1}$ (see below) and $a$ = 10$^{-5}$ cm implying that $Q_{abs}$ = 0.13, then $V_{gr}$ ${\approx}$ 10.0 km s$^{-1}$.

We measure at 1{\arcsec} from the star that
$I_{\nu}$(11.7 ${\mu}$m) ranges from 140 to 280 Jy arcsec$^{-2}$.  Adopting a ``typical" value of 200 Jy arcsec$^{-2}$, and with $V_{gr}$ = 10.0 km s$^{-1}$, ${\chi}_{\nu}$ (11.7 ${\mu}$m) = 2000 cm$^{2}$ g$^{-1}$, and $f_{corr}$ ${\approx}$ 2.0, then, from equation (4), 
${\dot{M}_{gr}}$ ${\approx}$  1.6 ${\times}$ 10$^{-9}$ M$_{\odot}$ yr$^{-1}$.

We  test this  model by comparing its predictions with observations.  As shown in Figure 5, the model and the spatial variations of the observed intensity at both 11.7 ${\mu}$m agree reasonably well with each other.  However, as shown in Figure 6, the agreement between the model and the observations at 17.9 ${\mu}$m do not
agree especially well.  As discussed above, the data indicated a temperature
at 1{\arcsec} of 400 K while the model predicted a temperature of 280 K.
Since the intensities at 17.9 ${\mu}$m may be systematically in error
because of the difficulty of calibrating our observations,   we scale  all the 17.9 ${\mu}$m intensities upward by a factor of 2 and plot these results in Figure 6 as well.  In this
case, the model agrees reasonably well  with the observations.      

In addition to determining the dust loss rate, we would like to estimate the gas loss
rate.  The most common tracer of the amount of circumstellar gas is CO.  However, 
the circumstellar envelope around L$_{2}$ Pup is unusual.  Kerschbaum \& Olofsson (1999) report that the integrated intensity in the  CO (2-1) lines compared to that in the CO (1-0) line is 12, substantially larger than the value for this ratio measured for other semiregular stars  and also larger than can be produced by any standard model
for thermal excitation of this molecule (see Kahane \& Jura 1994).  
Since we  have only an uncertain measure of the amount of circumstellar gas, we adopt  ${\dot M}_{ga}/{\dot M}_{gr}$ ${\approx}$ 200, consistent with the inferred evolutionary status of the star as a member of the thick disk.  With a dust
loss rate of 1.6 ${\times}$ 10$^{-9}$ M$_{\odot}$ yr$^{-1}$, then the total mass
loss rate is ${\sim}$3 ${\times}$ 10$^{-7}$ M$_{\odot}$ yr$^{-1}$.

\section{PULSATIONAL MODE}

L$_{2}$ Pup is a variable star, and the models of Winters et al. (2000) are consistent with our inferred mass loss rate. Here, we note that the energy carried away by the wind, $(1/2)\,{\dot {M}_{ga}}\,V_{ga}^{2}$, is about
1 ${\times}$ 10$^{30}$ erg s$^{-1}$. Wood \& Karovska (2000) observed the Mg II ultraviolet emission from L$_{2}$ Pup during different phases of its pulsational cycle.  This line is excited by shocks caused by the pulsations, and while variable, a typical flux
at the Earth is 5 ${\times}$ 10$^{-13}$ erg s$^{-1}$.  Therefore, shocks
in the atmosphere radiate at least 2 ${\times}$ 10$^{29}$ erg s$^{-1}$, and  thus there appears to be enough kinetic energy in the pulsations  to drive the outflow, lending support to the using the models of Winters et al. (2000)
for explaining the wind.

Red giants often exhibit asymmetrical mass loss (see, for example,
Monnier, Tuthill \& Danchi 2000, Uitenbroek, Dupree \& Gililand 1998) whose origin is
not well understood.  It
is possible that non-radial pulsations at least partly
contribute to anisotropic mass loss, and this may be occuring around L$_{2}$ Pup.  
Here, we argue that 
 the hypothesis of non-radial pulsations is supported from observations of the time-variability of L$_{2}$ Pup at different infrared wavelengths. The basic idea is that the time variations in the flux from the circumstellar shell do not track very well the time variations in the flux emitted by the photosphere.  One possible explanation for
this difference is that the star undergoes non-radial pulsations.

A comprehensive data base of infrared emission from bright stars was acquired 
with the DIRBE instrument on the COBE satellite where fluxes at 9 different infrared bands were measured, sometimes more often than daily,  over a 300 day period.  We have used the standard web-based tool to extract these data for L$_{2}$ Pup, and
the results for 7 bands are shown in Figure 7.  It can be seen that the source is substantially more variable at ${\lambda}$ ${\leq}$ 3.5 ${\mu}$m, the light emitted by the photosphere, than at ${\lambda}$ ${\geq}$ 4.9 ${\mu}$m, the  light  emitted by the circumstellar dust. To illustrate the effect somewhat differently, we show in Figure 8, a plot
of the deviation from the mean value of the DIRBE fluxes at three wavelengths, 
1.25 ${\mu}$m, 2.2 ${\mu}$m and 12 ${\mu}$m.  The fluxes at 1.25 ${\mu}$m and
2.2 ${\mu}$m track each other, while the flux at 12 ${\mu}$m displays a
distinctly different time variation.  

The data in Figure 7 can be compared to
a simple model  for a spherically symmetric envelope where 
the specific luminosity of the dust, $L_{\nu}$, from the optically thin envelope is:
\begin{equation}
L_{\nu}\;{\approx}\;{\int}^{\infty}_{0}\,{\chi}_{\nu}(4{\pi}\,\,B_{\nu}[T_{gr}])\,(\frac{{\dot M}_{gr}}{V_{gr}})\,dR
\end{equation}
If the temperature variation of the circumstellar dust is given by equation (1), then we can show from equation (6) with a simple substitution of variables that:
\begin{equation}
 L_{\nu}\;{\propto}\;T_{*}^{2.5}\,R_{*} \;{\propto}\;(L_{*}\,T_{*})^{1/2}
\end{equation}  

Consider now the observations of L$_{2}$ Pup shown in Figures 7 and 8.   From the model atmospheres computed by Houdashelt et al. (2000), near $T_{eff}$ = 3400 K, we
expect that 
\begin{equation}
\frac{{\Delta}T}{T}\;{\approx}\;0.4\frac{{\Delta}r_{nir}}{r_{nir}}
\end{equation}
where $r_{nir}$ denotes the flux ratio, F$_{\nu}$(1.25 ${\mu}$m)/F$_{\nu}$(2.2 ${\mu}$m).
 For example, from  day 120 to day 190, the flux ratio, $r_{nir}$  increases by a factor of about 1.08 which can be explained by an increase in the temperature by 1.03. Although the variation in radius during this phase is not certain, in view of the total increase in the observed values of F$_{\nu}$(1.25 ${\mu}$m) and F$_{\nu}$(2.2 ${\mu}$m) as well as in $r_{nir}$, it is probable that during this time interval, the radius slightly increased as well.  Consequently, using equation (7), we expect from the inferred increase in $T_{*}$ that F$_{\nu}$(12 ${\mu}$m) should have risen by at least  8\%. In fact, as can be seen in Figure 8, F$_{\nu}$(12 ${\mu}$m) at day 190 is within 1\% of its value at day 120.  Furthermore, the time variation of the flux at 12 ${\mu}$m is
different from that at 2.2 ${\mu}$m.  Therefore, a model with radial pulsations
is not supported by these DIRBE observations.  One
way to understand these data is to presume that the star undergoes
nonradial pulsations.  In this case, F$_{\nu}$(12 ${\mu}$m)  measures the luminosity of the entire star while F$_{\nu}$(2,2 ${\mu}$m) only measures emission from the hemisphere of the star that faces the Earth.  A test of this model is that the resolved dust blobs shown in Figures 2 and 3 should display unsynchronized variations.

Additional evidence for nonradial pulsations comes from   
  the  marked time variation of the position angle of the net polarization of the star (Magalhaes et al. 1986).  These data can be explained if the
circumstellar blobs are differentially illuminated by a non-spherical
time-varying photosphere as would occur during nonradial pulsations.    

The DIRBE data can be used to study time-variations from other pulsating red giants besides L$_{2}$ Pup.  We show, for example,  in Figures 9 and 10, the DIRBE fluxes and their deviations from their mean values, for R Cas, a mass-losing infrared-bright Mira red giant with a period of 430 days which lies at a distance of about 110 pc from the Sun.  This star exhibits time variations at both near infrared and mid-infrared wavelengths which follow each other as would be expected from radial pulsations.

\section{DISCUSSION}

With $\dot{M_{ga}}$ ${\approx}$ 3 ${\times}$ 10$^{-7}$ M$_{\odot}$ yr$^{-1}$, V$_{gas}$ ${\approx}$ 3.5 km s$^{-1}$ and $L$ ${\approx}$ 1500 L$_{\odot}$,  the circumstellar envelope around L$_{2}$ Pup is well described
by the models  computed by Winters et al. (2000) where radiation
pressure on the grains is relatively unimportant in the dynamics. To assess this hypothesis, consider
 ${\alpha}$ which denotes the ratio of the outward radiation pressure on the circumstellar envelope (gas plus dust) compared to the inward force of gravity.  We may write (Winters et al. 2000)
\begin{equation}
{\alpha}\;=\;(\frac{{\dot{M_{gr}}V_{ga}}}{{\dot{M_{ga}}}V_{gr}})\;(\frac{L_{*}\,{\chi}}{4\,{\pi}\,c\,G\,M_{*}})
\end{equation}
With the parameters given above,  then for the envelope around L$_{2}$ Pup,  ${\alpha}$ = 0.6 and radiation pressure does not dominate the circumstellar dynamics.  

Besides L$_{2}$ Pup, there are a few other evolved red giant stars where the measured circumstellar gas exhibits velocity widths smaller than 5 km s$^{-1}$.  Jura \& Kahane (1999) have proposed that some of these stars, such as the Red Rectangle and AC Her which are known close binaries (van Winckel et al. 1998), possess orbiting molecular reservoirs.  They hypothesize that  the orbiting material may be produced
by interactions of the secondary star with the envelope ejected by the mass-losing star.  There is no evidence for a companion around L$_{2}$ Pup.  
However, at 61 AU from the star, the region where our infrared data indicates the presence of a large amount of circumstellar dust, the escape velocity is 5.4 km s$^{-1}$ which is larger than the inferred gas outflow speed.  If L$_{2}$ Pup has an undetected low-mass companion, the interaction of the wind with this star might produce some gravitationally bound matter in a circumbinary disk. 

An red giant  star may be spun-up by engulfing a companion, and Soker (2000) has proposed that such stars with relatively high rotation rates
may sometimes eject mass in a very flattened slow-moving outflow.  However, there is no reported evidence that L$_{2}$ Pup is rotating particularly rapidly.  Also, although this picture of a flattened outflow may partially explain the observation around L$_{2}$ Pup of an apparent  axis of symmetry at position angle 135$^{\circ}$, the presence of a distinct blob at position angle 225$^{\circ}$  is not naturally explained with this model.  

The current mass-loss phase of L$_{2}$ Pup  may be important in its evolution.
If L$_{2}$ Pup persists in  losing mass at ${\sim}$ 3 ${\times}$ 10$^{-7}$ M$_{\odot}$ yr$^{-1}$ for  5 ${\times}$ 10$^{5}$ yr, consistent with the models for stellar evolution computed by Lattanzio (1991) of the AGB or by Girardi et al. (2000) for the tip of the Red Giant Branch, then it will
lose ${\sim}$0.15 M$_{\odot}$ or ${\sim}$15\% of its initial main sequence mass.
This estimate of the lifetime of L$_{2}$ Pup in its current phase is completely theoretical
since it  takes the dust only ${\sim}$ 100 yr  to reach the projected
separation of 2{\arcsec} from the star seen in our infrared images. 

 L$_{2}$ Pup  may serve as
a test of the proposed mass loss rate from  red giants.  This is important  because it is not possible to compute with confidence the mass loss
rate from our current understanding of stellar evolution  (see, for example, Catelan 2000, Schroder \& Sedlmayr 2001).  Although recognizing that there
are many uncertainties, Catelan
(2000) has proposed that the mass loss rate from red giant stars can be parameterized  by
the expression
\begin{equation}
\dot{M}_{ga}\; =\; 8.5 {\times} 10^{-10} \left(\frac{L_{*}}{gR_{*}}\right)^{1.4}\;\;M_{\odot}\,yr^{-1}
\end{equation}
where $g$ is the acceleration of the star (cgs), and $L_{*}$ and $R_{*}$ are in solar units.  In the case of L$_{2}$ Pup, this formula predicts a mass loss
rate of 1.1 ${\times}$ 10$^{-8}$ M$_{\odot}$ yr$^{-1}$, a factor of 30 less
than we infer.  It could be that the formula of Catelan (2000) fails to describe
adequately the mass loss from L$_{2}$ Pup because it does not distinguish
between pulsating and non-pulsating stars. 

Some  red giants in 47 Tuc with roughly the same luminosity as
L$_{2}$ Pup do not exhibit any 12 ${\mu}$m excess (Ramdani \& Jorissen 2001). A similar result  is found for red giants in the neighborhood of the Sun  (Jura et al. 2001).   If these stars are losing mass, it is without the production of large amounts of dust.     In the models of Winters et al. (2000), the dust is not necessary to drive the mass
loss, and thus there might be red giants with large mass loss rates and little dust
formation.   Catelan's formula possibly applies to non-pulsating red giants with luminosities near
1000 L$_{\odot}$.   

\section{CONCLUSIONS}
We have obtained mid-IR images of the dust around L$_{2}$ Pup.
We find the following:
\begin{itemize}
\item{We estimate a total mass loss rate of ${\sim}$3 ${\times}$ 10$^{-7}$ M$_{\odot}$ yr$^{-1}$. During the current  phase of its evolution, L$_{2}$ Pup may lose ${\sim}$15\% of its initial main sequence mass.}
\item{ Non-radial pulsations may at least in part account for the observed asymmetric mass loss, time variations of the infrared fluxes, and time
variations of the position angle of the optical polarization.}
\item{L$_{2}$ Pup might serve as the prototype of an outflow driven by stellar pulsations with radiation pressure on dust
being relatively unimportant, as in  models described
by Winters et al. (2000).}

This work has been partly supported by NASA.
\end{itemize}
   
\newpage
\begin{center}
{\bf FIGURE CAPTIONS}
\end{center}
Fig. 1.  Histogram of outflow velocity measured from the profile of the circumstellar CO emission for a sample of semiregular and irregular variables (Kerschbaum \& Olofsson 1999).  In the case of several measurements of the line profile, we adopt the average value.  In the case of profiles where both a narrow and broad component were identified, we consider only the broad component.  The location of
L$_{2}$ Pup in this histogram is indicated.
\\
\\
Fig. 2.  The 11.7 ${\mu}$m image of L$_{2}$ Pup.  North is up and East
is to the left.  The contour levels (Jy arcsec$^{-2}$) are shown in the color bar.
\\
\\
Fig. 3.  The 17.9 ${\mu}$m image of  L$_{2}$ Pup with the same conventions as Fig. 2.  The very faint blob located at 2${\arcsec}$ east of the star is an artifact of
the LWS filter at this wavelength.  
\\
\\
Fig. 4.  The 11.7 ${\mu}$m image of Capella with the same conventions
as Fig. 2.
\\
\\
Fig. 5. The   measurements of $I_{\nu}$(11.7 ${\mu}$m) at offsets ${\geq}$ 0{\farcs}64 from the star at position angle 0$^{\circ}$ (circles) and
120$^{\circ}$ (crosses). The solid line shows the predictions for $I_{\nu}$(11.7 ${\mu}$m) for the spherical model described in the text with ${\dot {M_{gr}}}$ = 1.6 ${\times}$ 10$^{-9}$ M$_{\odot}$ yr$^{-1}$, $V_{gr}$ = 10.0 km s$^{-1}$ and
$T_{gr}$ = 280 K at 61 AU from the star.    
\\
\\
Fig. 6.  The same as Figure 5 except that we compare data at 17.9 ${\mu}$m with the model shown as the solid line.  As described in the text, the calibration of the data is uncertain;  the red points show the same data
scaled upwards by a factor of 2.   
\\
\\ 
Fig 7.  The  fluxes acquired by the DIRBE instrument on COBE for L$_{2}$ Pup.  The labels refer to fluxes at different bands and are reported at different days during the COBE mission relative to January 1, 1990.
\\
\\
Fig. 8.  The flux ratios measured by DIRBE for L$_{2}$ Pup relative to the mean value at different times for 1.25 ${\mu}$m,  2.2 ${\mu}$m and 12 ${\mu}$m.  
The curves at 1.25 ${\mu}$m and 2.2 ${\mu}$m follow each other and track the emission from the photosphere while the 12 ${\mu}$m emission is mainly
produced by circumstellar dust.    
\\
\\
Fig. 9.  The fluxes acquired by the DIRBE instrument on COBE for R Cas with
the same conventions as in Figure 7.  The time variations of the fluxes at 12 ${\mu}$m and 25 ${\mu}$m follow
those at shorter wavelengths.
\\
\\
Fig. 10. The flux ratios measured by DIRBE for  R Cas relative to the mean value at different times for 1.25 ${\mu}$m, 2.2 ${\mu}$m and 12 ${\mu}$m. 
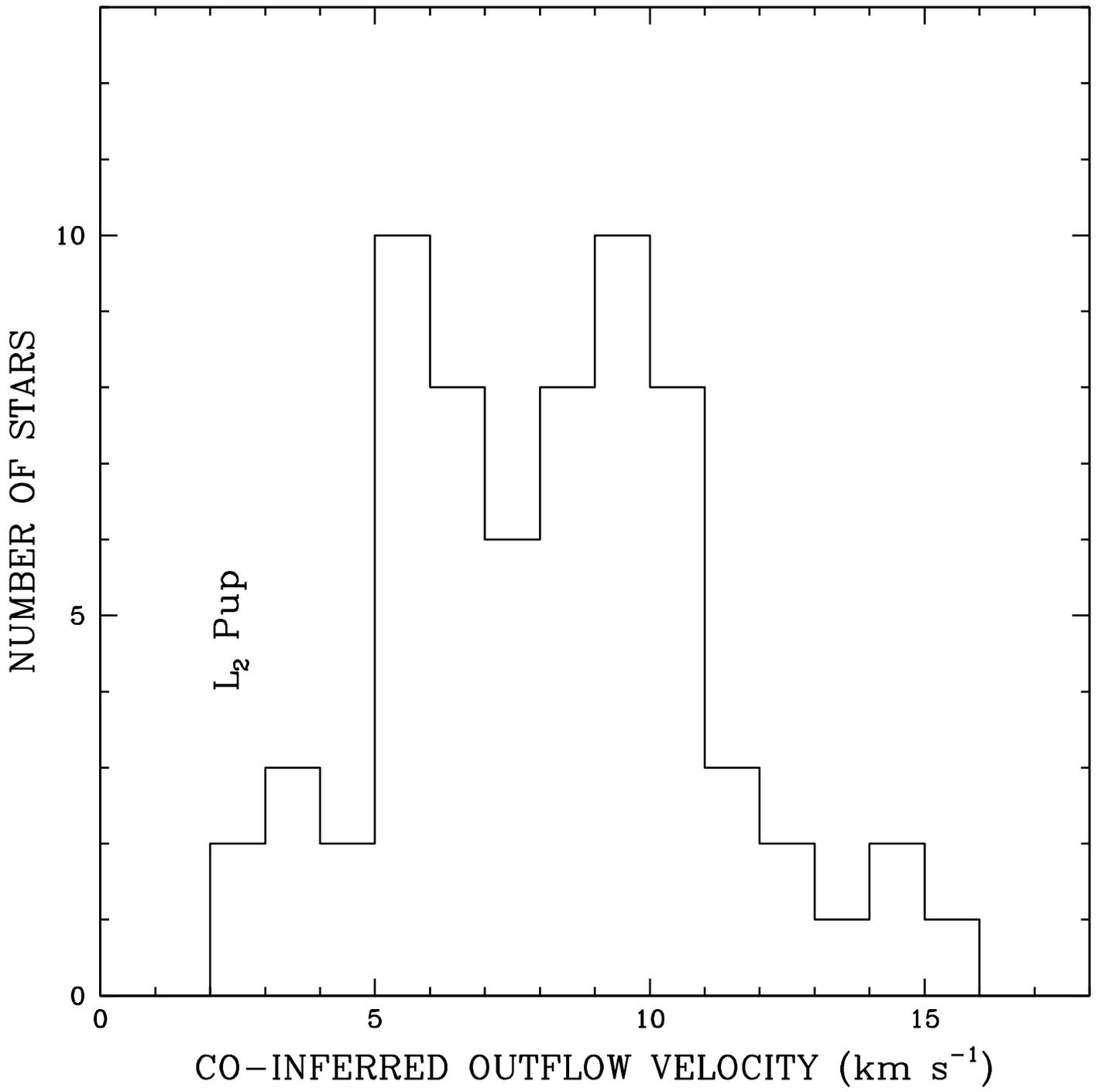
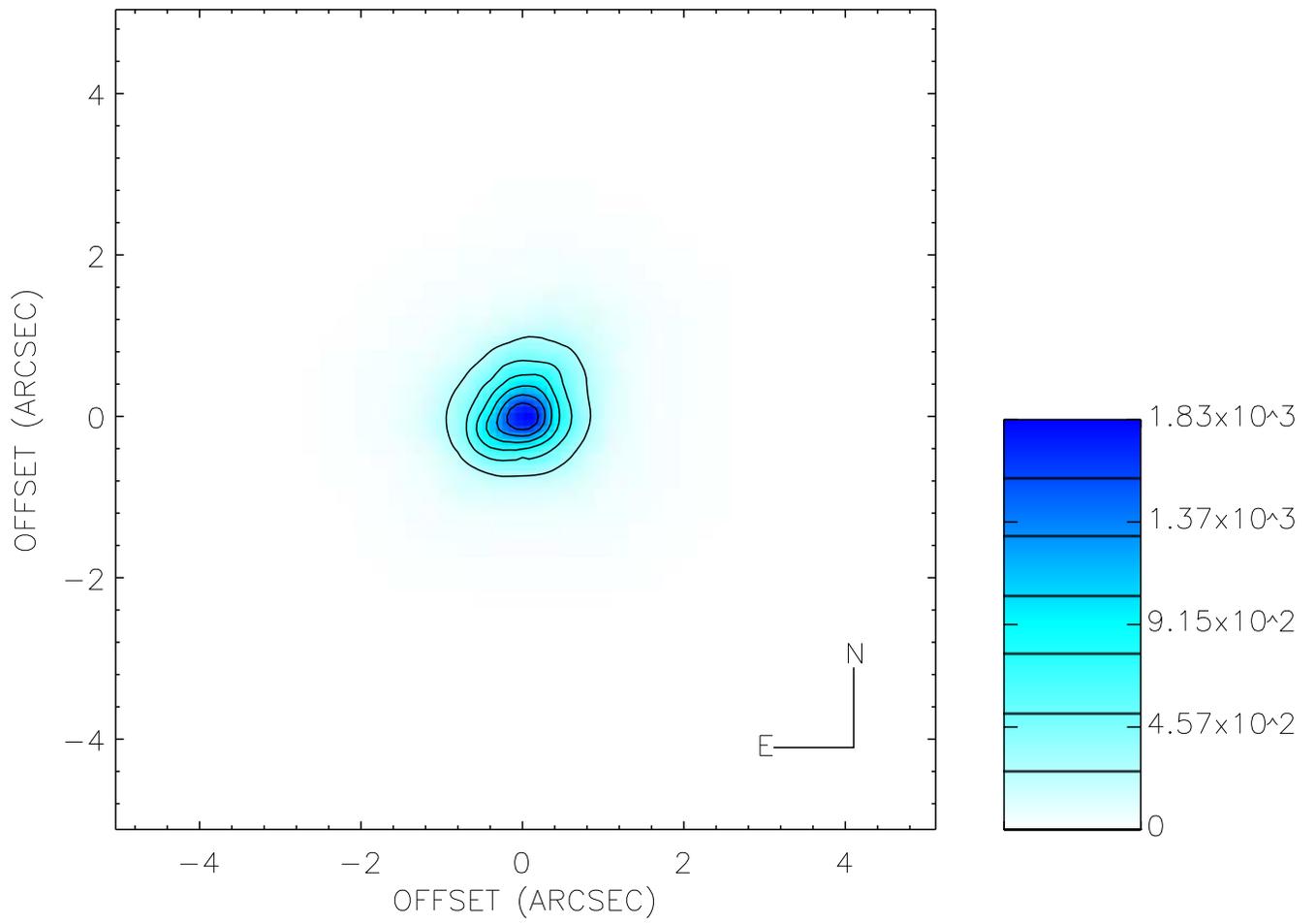
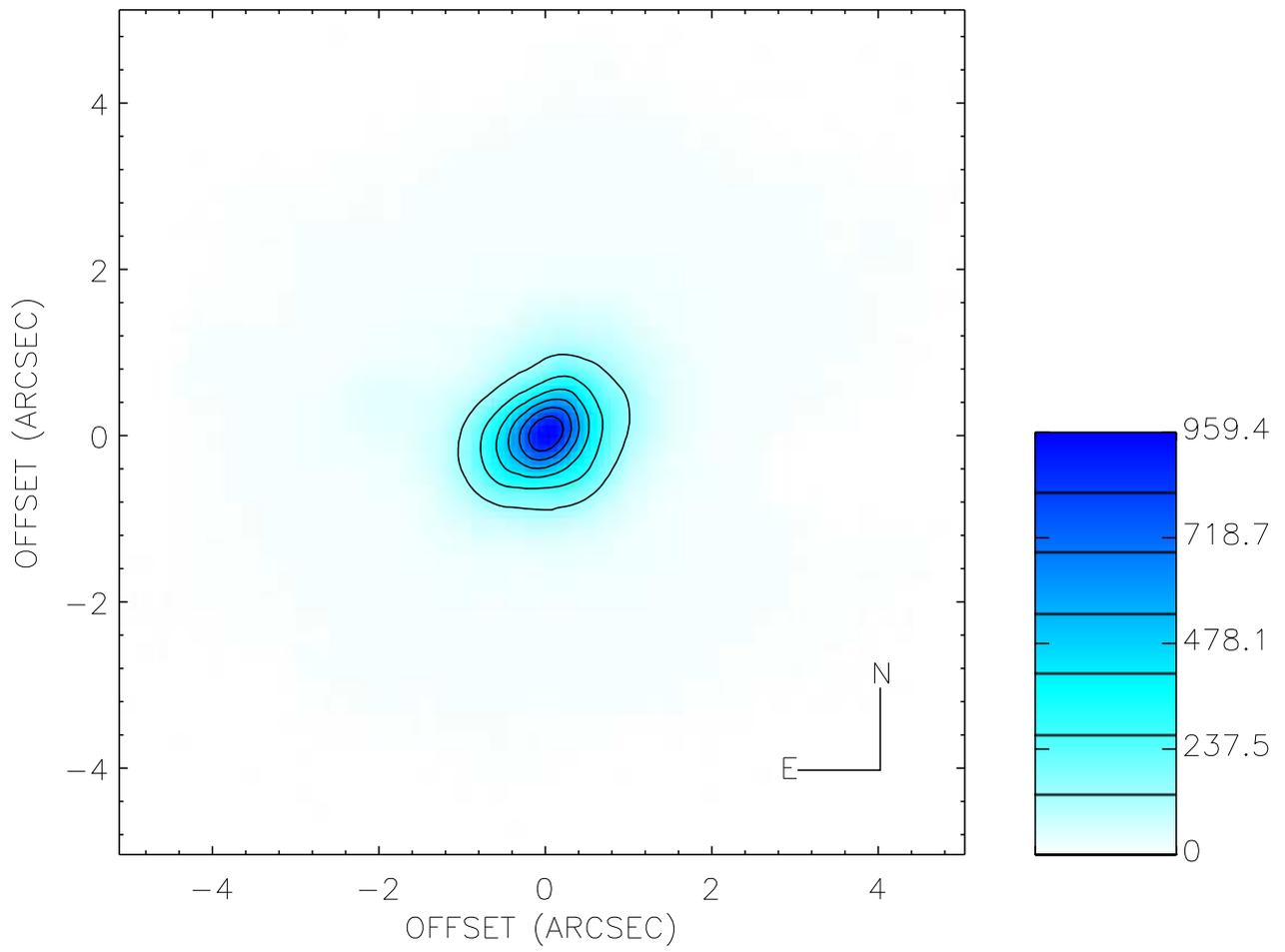
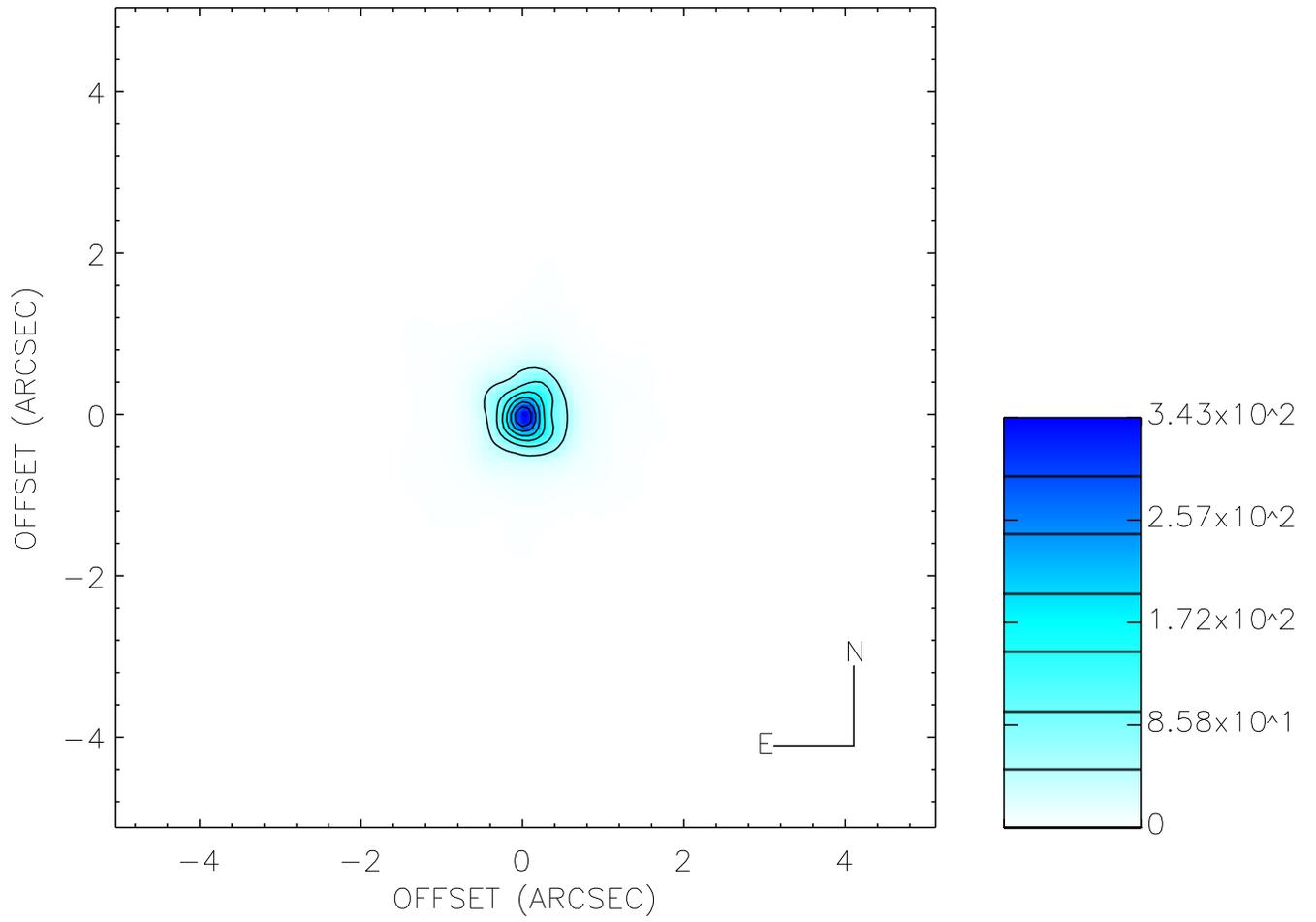
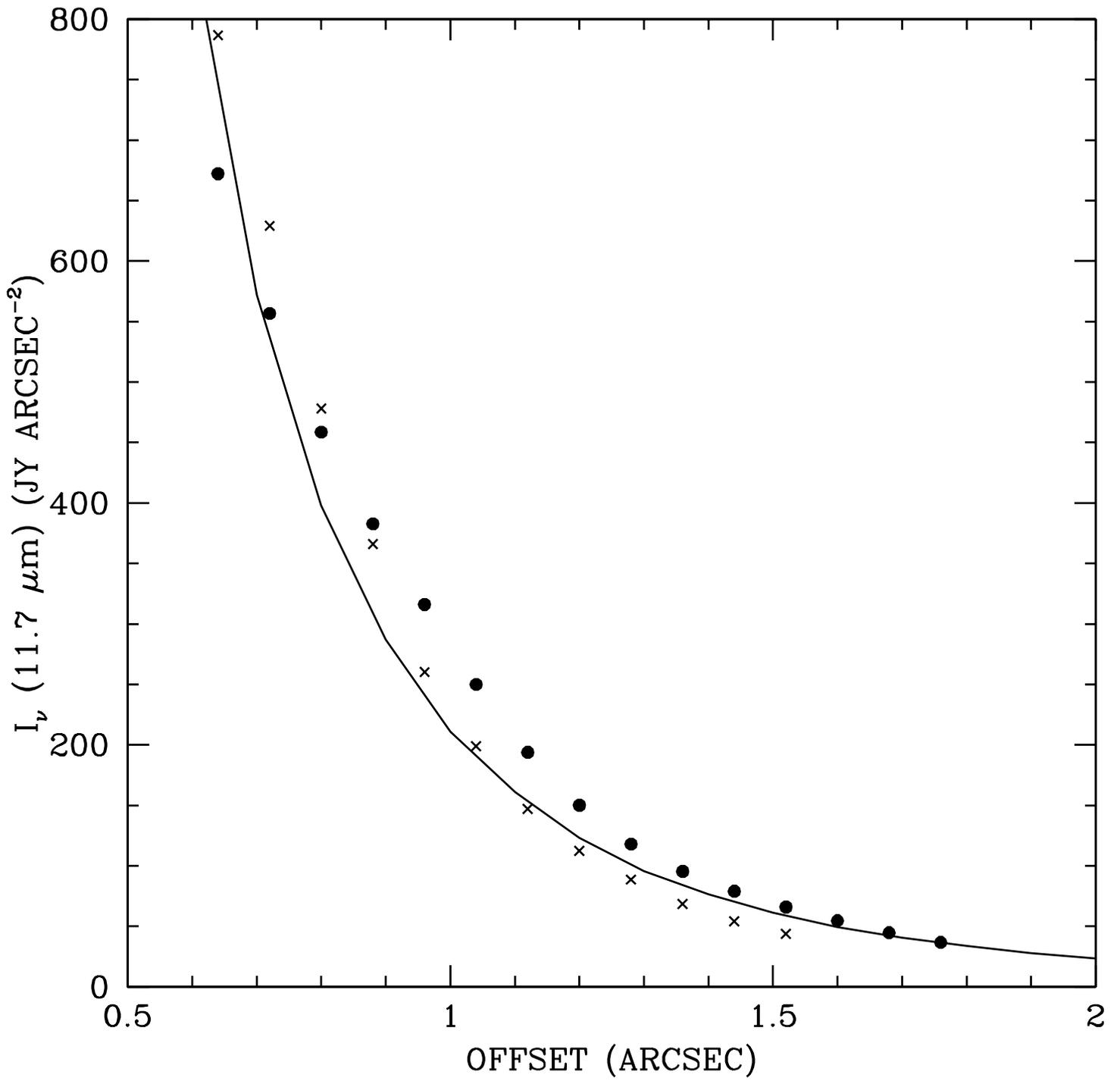
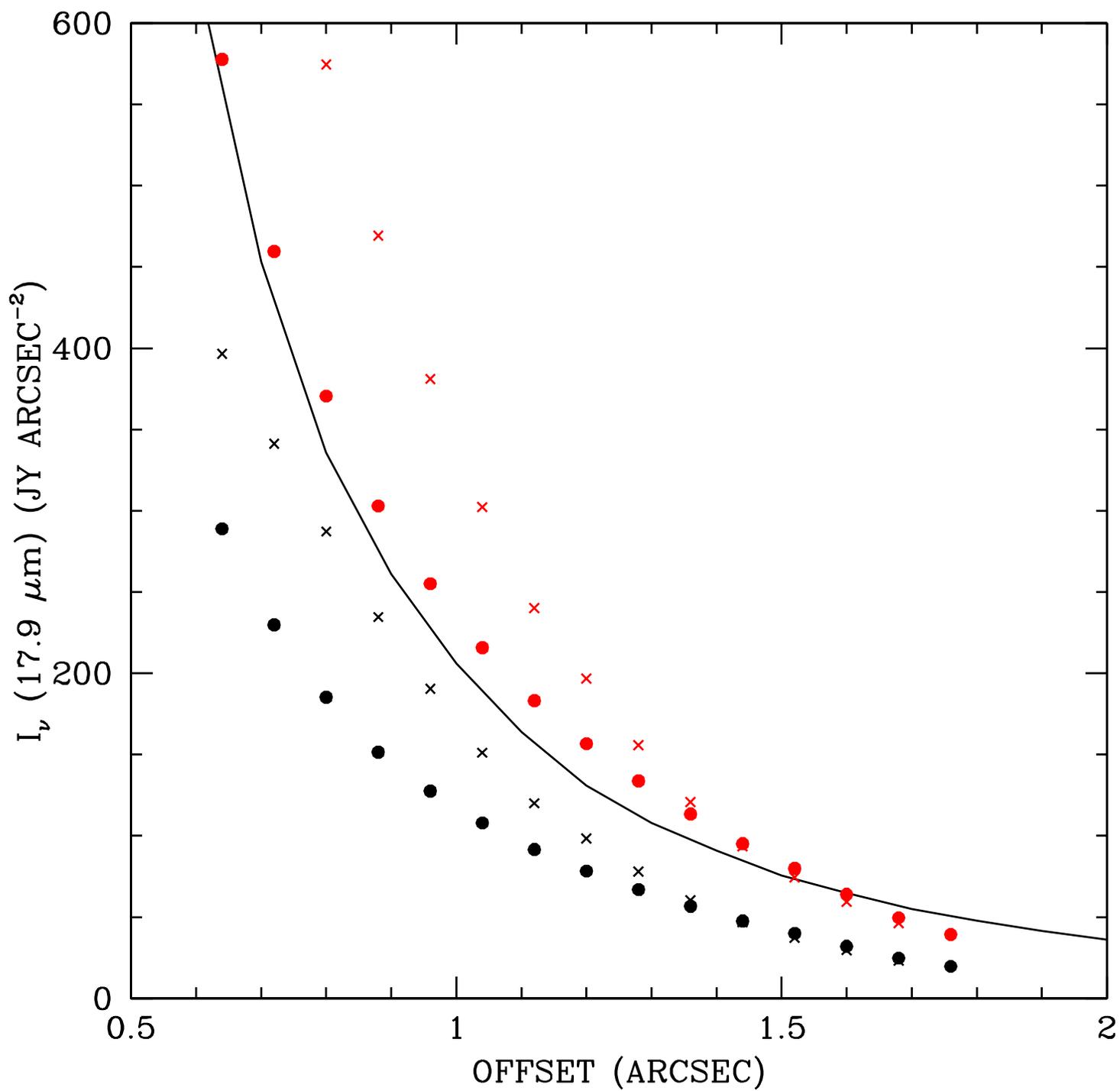
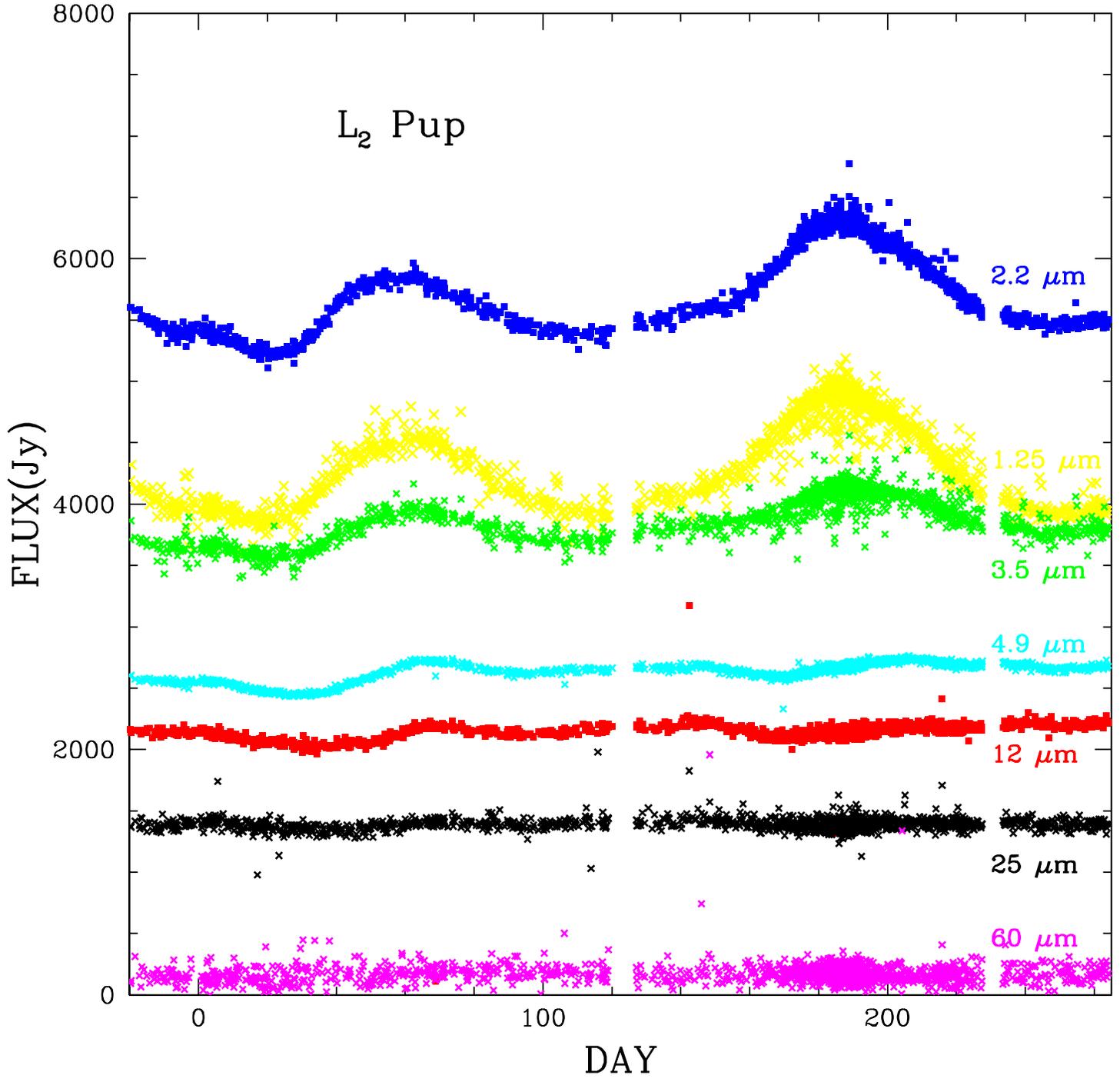
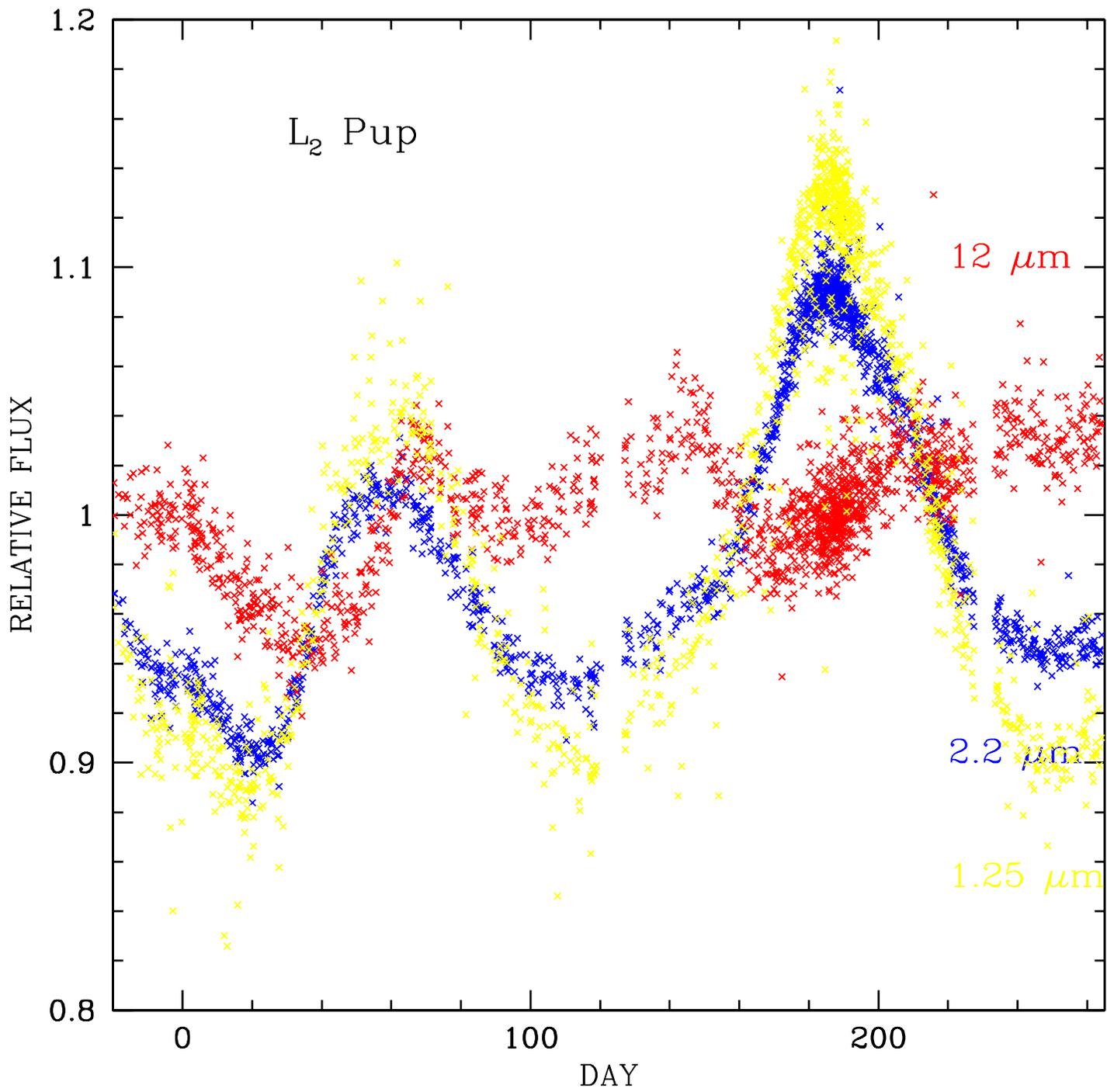
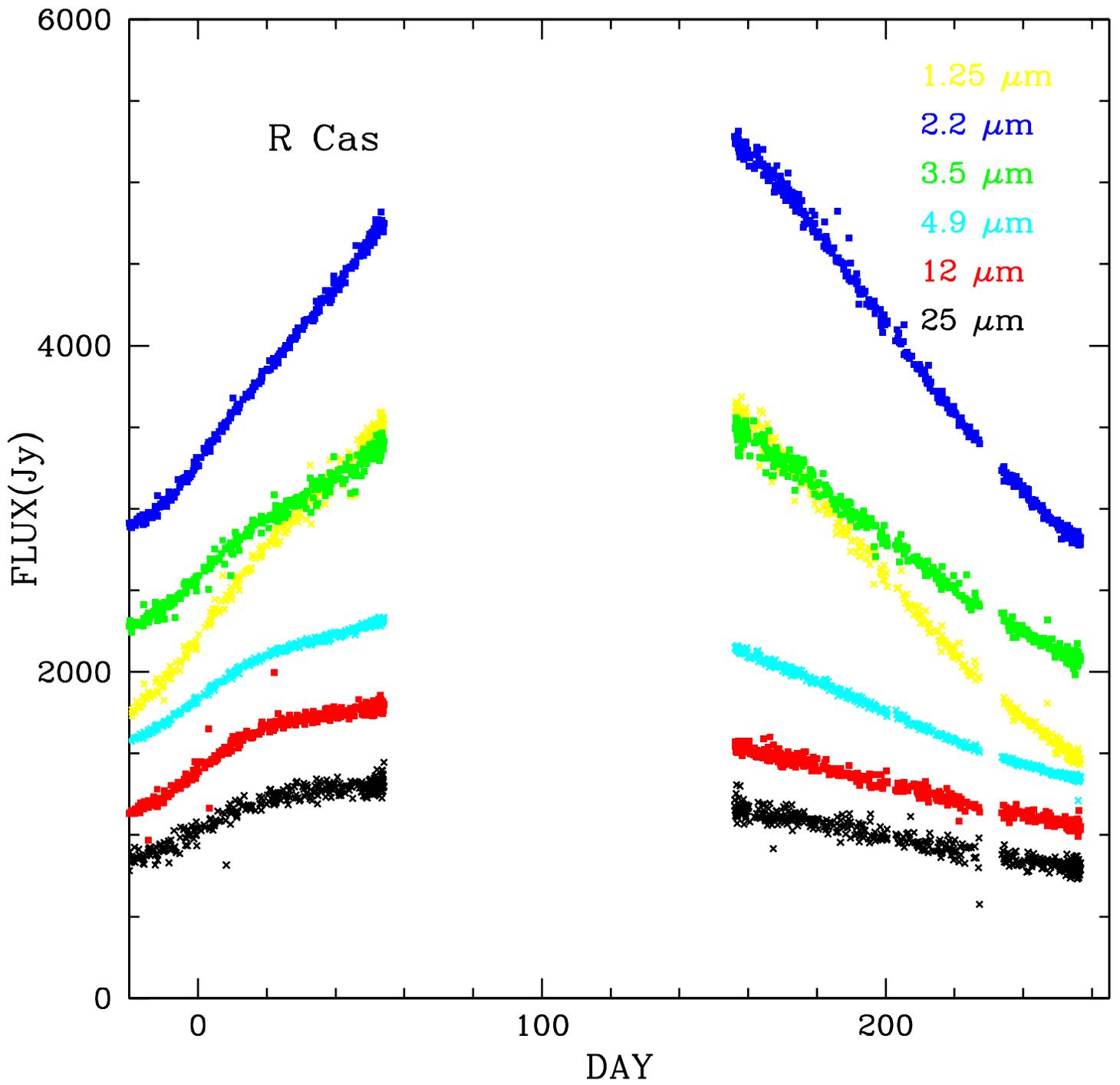
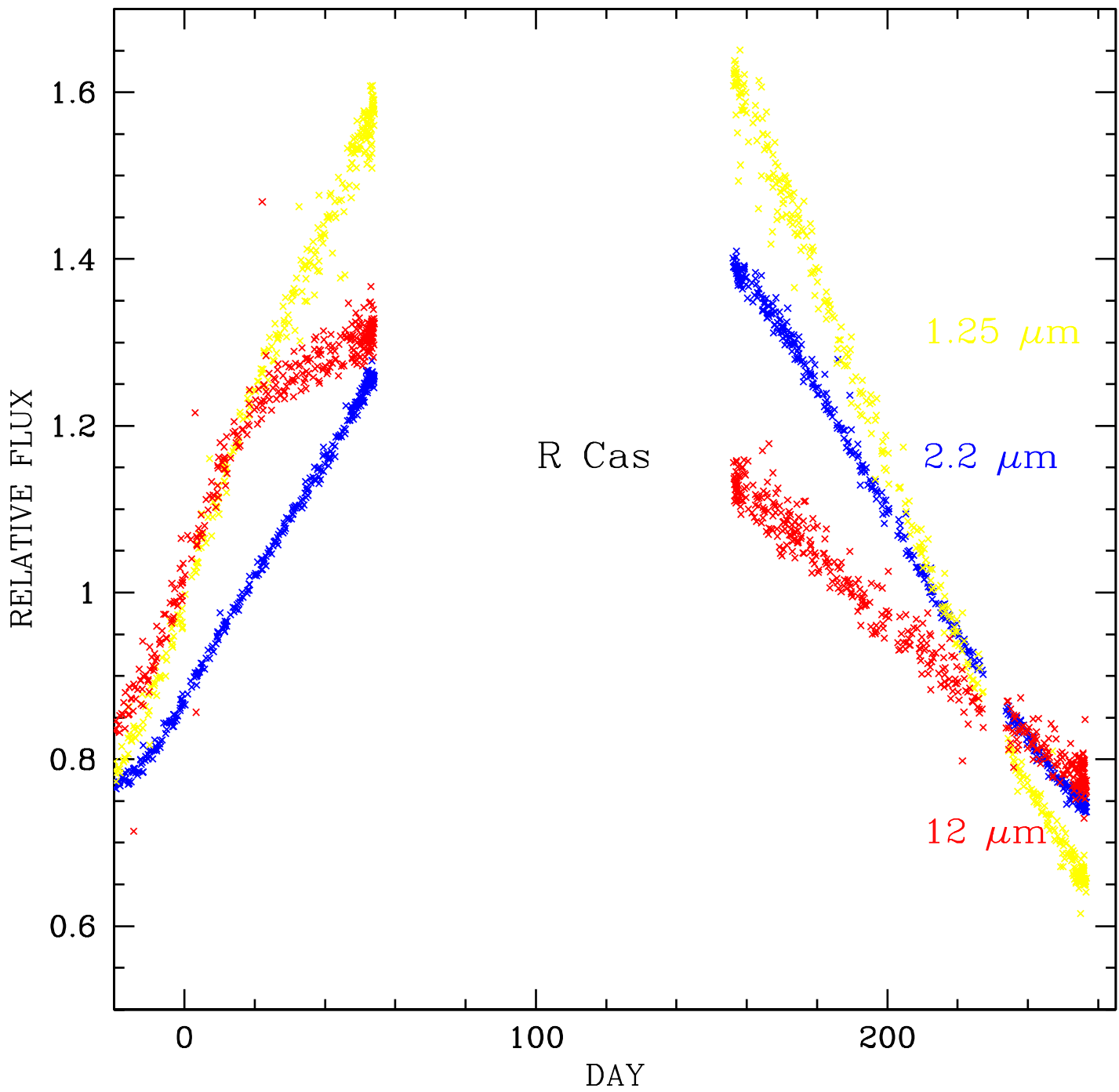
\end{document}